\documentclass[prl,amsmath,amssymb]{revtex4}
\usepackage{graphicx}
\usepackage{bm}
\usepackage{dcolumn}
\usepackage{subfigure}
\usepackage{psfrag}
\newcommand{\vk}{{\vec k}}
\newcommand{\be}{\begin{equation}}
\newcommand{\ee}{\end{equation}}
\newcommand{\p}{\partial}
\begin{document}

\title{Graphene as an electronic membrane}

\author{ Eun-Ah Kim$^1$, and A. H. Castro Neto$^2$,}

\affiliation{$^1$ Stanford Institute for Theoretical Physics and 
Department of Physics, Stanford University, Stanford, California 94305}
\affiliation{$^2$Department of Physics, Boston University, 590 
Commonwealth Avenue, Boston, MA 02215}
\email[Correspondence should be send to:]{neto@bu.edu}

\date{\today}

\maketitle

\textbf{Experiments are finally revealing intricate facts about graphene which go beyond the 
ideal picture of relativistic Dirac fermions in pristine two dimensional (2D) space, two years 
after its first isolation\cite{Netal04_short,geim_review}. While observations of rippling 
\cite{meyer07_short,kim_STM,Ishigami_STM}  added another dimension to the richness of the physics 
of graphene, scanning single electron transistor images displayed prevalent charge inhomogeneity \cite{yacoby}. 
The importance of understanding these non-ideal aspects cannot be overstated both from the fundamental 
research interest since graphene is a unique arena for their interplay, and from the device applications 
interest since the quality control is a key to applications. We investigate the membrane aspect of 
graphene and its impact on the electronic properties. We show that curvature generates spatially 
varying electrochemical potential. Further we show that the charge inhomogeneity in turn stabilizes ripple 
formation.}

Since its unexpected isolation \cite{Netal04_short}, free standing graphene, a single atomic layer of 
carbon atoms forming a 2D honeycomb lattice,  has risen as an intriguing and promising metamaterial. 
Not only the charming notion of manipulating relativistic fermions on a table top 
but also the potential of the graphene-based electronics, is propelling the current enthusiasm
\cite{geim_review,pw}. However, for the graphene-based electronics it is vital to understand the 
interplay and connection among observed real material aspects. 

The observations of ripples in suspended graphene using transmission electron microscopy (TEM) 
\cite{meyer07_short} and in graphene on SiO$_2$ substrate using scanning tunneling microscopy (STM) 
\cite {kim_STM, Ishigami_STM} summon the membrane aspect of graphene  to the foreground. Statistical 
mechanics of membranes has long been an important branch of soft condensed matter physics\cite{nelson_book},  
with its application to biological systems being one of its driving forces. However, it has been irrelevant 
for the study of conventional 2D electronic systems which are buried in the semiconductor heterojunction 
structure. Graphene being a single layer of carbon atoms that can be gated, it forms the first example of 
an electronic membrane that is subject to direct probes. 

Electron-hole puddles in graphene imaged by scanning single electron transistor (SET) \cite{yacoby} suggest 
that such charge inhomogeneity should play an important role in limiting the transport characteristics of 
graphene. One cause of such charge inhomogeneity could be remote charged impurities in the substrate as it 
was noted in Refs.\cite{hwang,nomura}.  By building an effective theory based on microscopic calculations, 
we find the corrugations to generate inhomogeneous electrochemical potential directly on the graphene membrane; 
thus identify corrugations as another cause of charge inhomogeneity.  We predict control over irregular 
corrugations to improve the transport properties of graphene greatly. At present, the technology for 
stabilizing perfectly flat graphene is not available.  One possibility is to epitaxially grow monolayer 
graphene in a controlled manner. While current epitaxial growth technique only yields a few layer of 
graphene with the bottom layer forming the dominant channel for transport\cite{epitaxy},  
likely the status will improve in a near future. 

\begin{figure}[b]
\subfigure[]{
\includegraphics[width=.4\textwidth]{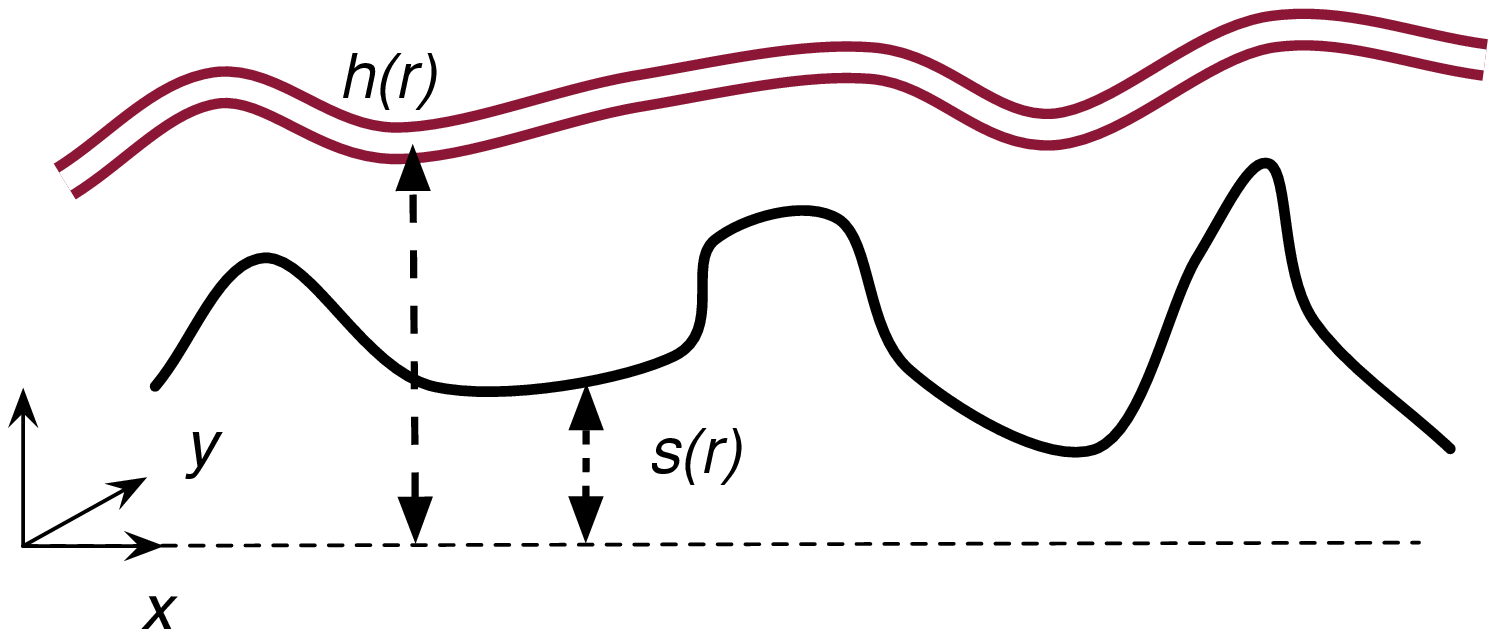}}
\includegraphics[width=.2\textwidth]{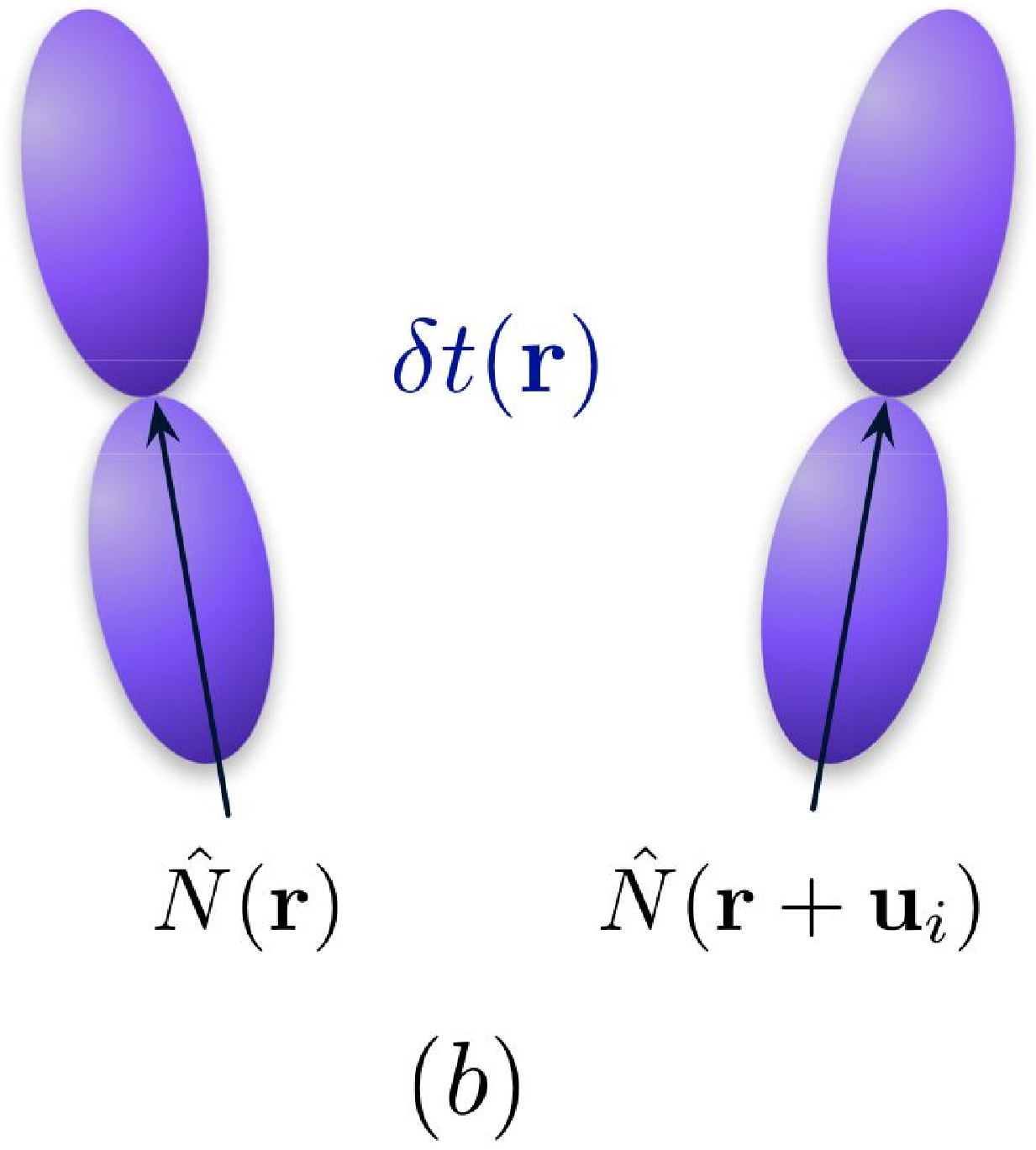}$\qquad$
\includegraphics[width=.2\textwidth]{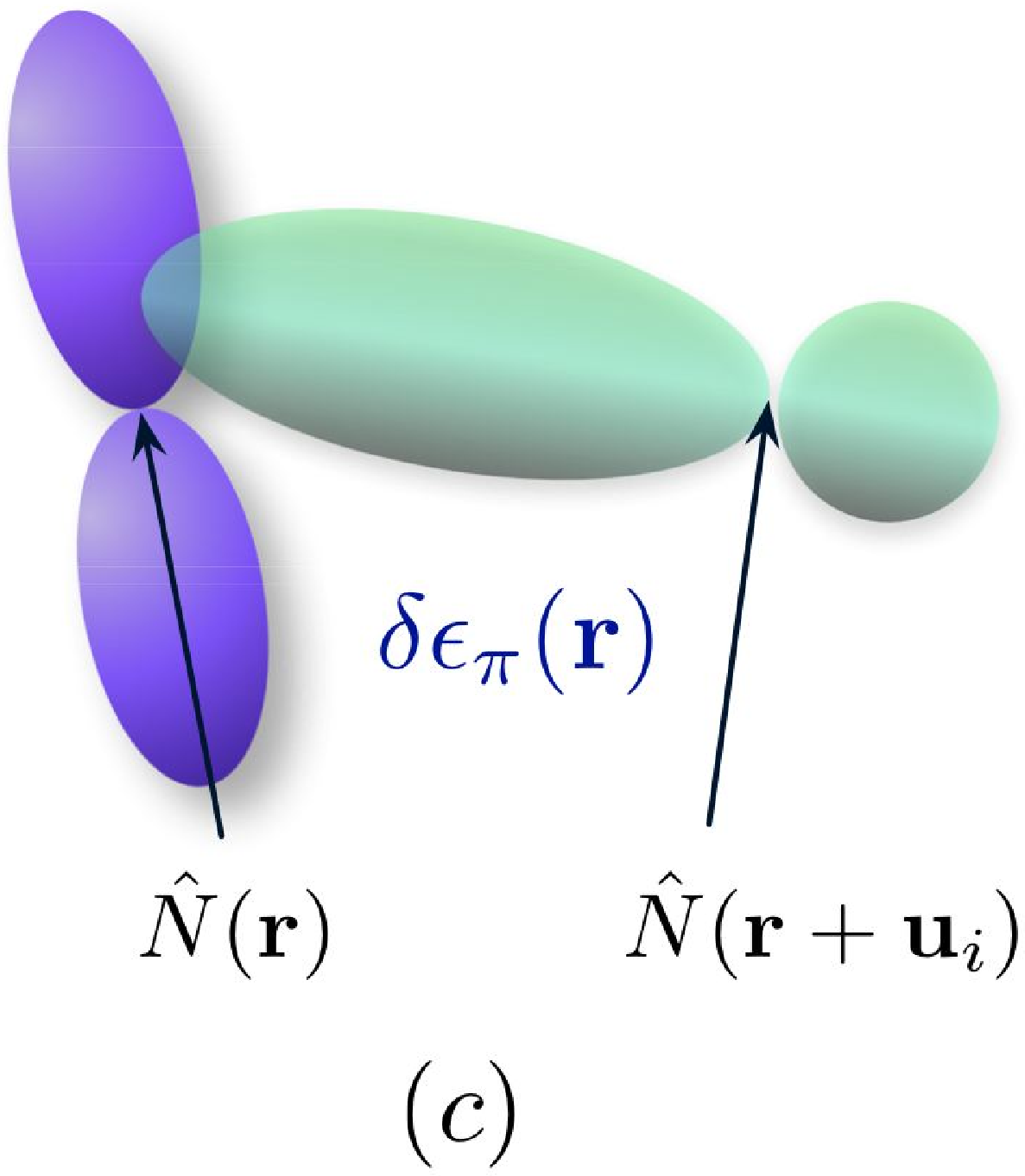}
\caption{Graphene as an electronic membrane. (a) The reference $\bf r$-plane is shown as a dashed line. The height of the curved graphene surface and the substrate surface measured from this plane, are denoted by $h({\bf r})$ and $s({\bf r})$ respectively. (b)The curvature dependent hopping integral which depends on the overlap $\hat{N}({\bf r})\cdot\hat{N}({\bf r}+{\bf u}_i)$ introduces local ``vector potential''. 
(c) The $\pi$-$\sigma$ rehybridzation due to the curvature which results in local electrochemical potential variation.
}
\label{fig:surface}
\end{figure}

For the energetics associated with corrugations at large length scales, we employ a continuum model frequently used to study 
the statistical behavior of a thin membrane.  For a pure membrane freely fluctuating in a solvent, the parametrization independent bending energy is\cite{free-mem}
 \begin{equation}
E_{bending}=\int ds_1 ds_2\sqrt{g}\left[\gamma+\frac{1}{2}\kappa(2H)^2+\bar{\kappa}K\right]
\label{eq:bending}
\end{equation}
where $(s_1,s_2)$ are cooridinates intrinsic to the membrane; $g$ is the determinant of the metric; $\gamma$ is the tension; $\kappa$ is the bending rigidity; $\bar{\kappa}$ is the Gaussian rigidity; $H$ and $K$ are respectively the mean and the Gaussian curvature at the given point. If $R_1$ and $R_2$ are the local principle radii of curvature $H\!=\!(R_1^{-1}\!+\!R_2^{-1})/2$ and  $K\!=\!(R_1R_2)^{-1}$. For graphene, the gaussian curvature term drops out of the total bending energy since $\int d\sigma^2\bar{\kappa}K =4\pi\bar{\kappa}\chi$  where $\chi$ is the Euler characteristic set by the topology of the membrane configuration and $\chi\!=\!0$ for asymptotically flat surface as is the case for graphene ($\chi\!=\!1$ for fullerene) \cite{maria}. $\kappa \approx 1$ eV has been measured recently in graphene
electro-mechanical resonators \cite{mceuen07_short}
in agreement with microscopic calculations \cite{lenosky_short,tu}. 

Typically a sheet of graphene is held in place for measurements either by a substrate or by a scaffold. Such fixation breaks translational symmetry in the direction perpendicular to the device plane. The effect of such symmetry breaking can be accounted for by a potential $V(h,s)$ in addition to Eq.~\eqref{eq:bending}, where $s({\bf r})$ is the height of the substrate and a Monge representation $(x,y,h({\bf r}))$ is used to parametrize the membrane configuration which is approximately parallel to the $(x,y)$-plane (see Fig.\ref{fig:surface}).  
$V(h,s)$ includes an external pressure, an attraction to the support such as Van der Waals attraction, and a repulsive force. We follow Ref.\cite{swain99} and approximate the attraction to be local and expand the potential to second-order. The total elastic free energy then becomes:
\begin{equation}
\mathcal{F}[h]=\int d{\bf r}\left[ \gamma+\frac{1}{2}\gamma(\nabla h)^2+\frac{1}{2}\kappa(\nabla^2 h)^2 +\frac{1}{2}v(h-h_0-s)^2\right]
\label{eq:harmonicF}
\end{equation}
where $v=\partial^2 V(h_0,s)/\partial h^2$ for equilibrium configuration $h_0$. Tensionless membranes ($\gamma\!=\!0$) floating in solution are subject to crumpling instability\cite{nelson_book} which makes the mean-field approach inapplicable in such situation. However, in the presence of a finite tension which stabilizes directional order and the potential which cause the membrane to conform, the equilibrium configuration of the membrane can be found by solving the Euler-Lagrange equation (the mean-field equation). We find that as a result of interplay between the membrane aspect and the electronic structure of graphene, charge inhomogeneity favors mean curvature fluctuation and vice versa.  

The effect of the finite curvature on the band structure can be understood within the Slater-Koster prescription. There are two
main effects: a) the change in the $\pi$-orbital energy due to $\pi$-$\sigma$ mixing induced by curvature, b) the change in the nearest neighbor hopping integral. The first provides the link between the curvature and the electrochemical potential variation, while the second 
introduces effective ``vector potential''. 

The low energy effective Hamiltonian %for the $\pi$ electrons
 in the vicinity of a Dirac point $\vec{K}\!=\!(4\pi/3\sqrt{3}a,0)$ can be written as 
\be
H = \int d^2r\, \Psi^{\dag}({\bf r}) \left\{ \vec{\sigma} \cdot
 \left[iv_F\nabla +{\vec A} 
+ \vec{{\cal A}}({\bf r}) \right] - \mu +
\Phi({\bf r})+V({\bf r})\right\} \Psi({\bf r})
\label{eq:DiracH}
\ee
where $v_F\equiv3ta/2$ for bare nearest neighbor hopping integral $t\!=\!2.7{\rm eV}$ and the nearest neighbor distance $a\!=\! 2.5{\rm \AA}$, ${\vec \sigma}\!=\!(\sigma_x,\sigma_y)$ are Pauli matrices in the sublattice basis $(a_{\vk}, b_{\vk})$. $ \vec{{\cal A}}({\bf r})$ and  $\Phi({\bf r})$ are curvature induced ``vector potential'' and electrochemical potential whose role is of our main interest, and ${\vec A}$ is the external electromagnetic field and $V$ is the  charged impurity potential. (The effect of the impurity potential has been investigated in Refs.\cite{hwang,nomura} and it depends on the impurity strength and the screening not known from first principles)  

The curvature causes a misalignment between $\pi$ orbitals and $\pi$-$\sigma$ rehybridzation between nearest neighbors.   The ``vector potential'' due to the misalignment induced curvature effect on nearest neighbor hopping is (see Fig.~\ref{fig:surface}(b))
\begin{align}
&\mathcal{A}_x({\bf r}) + i\mathcal{A}_y({\bf r})=
 - \sum_j\delta t_j({\bf r})e^{i\vec{u}_j\cdot\vec{K}}
 =-\sum_j\frac{\epsilon_{\pi\pi}}{2}(({\vec u}_j\cdot\nabla)\nabla h)^2 e^{i {\vec u}_j\cdot {\vec K}}\nonumber\\
&\mathcal{A}_x
=-\epsilon_{\pi\pi}\frac{3a^2}{8}[
(\p_x^2h)^2-(\p_y^2h)^2],\qquad \mathcal{A}_y=\epsilon_{\pi\pi}\frac{3a^2}{4}\left[ \p_{x,y}^2h \, (\p_x^2h+\p_y^2h)\right],
 \label{eq:A}
\end{align}
where ${\vec u}_i$'s are three nearest neighbor vectors $\vec{u}_1=a({\sqrt{3}}/{2},{1}/{2}),\,
\vec{u}_2=a(-{\sqrt{3}}/{2},{1}/{2}),\, \vec{u}_3=a(0,-1)$, and   
$\epsilon_{\pi\pi}\!=\!V_{pp\pi}/3 \!+\!V_{pp\sigma}/2\! 
=\! 2.89{\rm eV}$ (See Ref.\cite{walter} for the definition of parameters). Eq.\eqref{eq:A} can be understood as the following. 
Small misalignment of angle $\theta$ affects the coupling between neighboring $\pi$ orbitals  and the coupling $V_{pp\pi}(=t)$ becomes 
$(V_{pp\pi} \cos^2 \theta \!-\!\frac{3}{2}V_{pp\sigma} \sin^2 \theta)  
\approx V_{pp\pi}\!-\!(V_{pp\pi}\!+\!\frac{3}{2}V_{pp\sigma})\theta^2$ and $\theta^2\sim(({\vec u}_j\cdot\nabla)\nabla h)^2$ in the presence of local curvature. While the gauge fields \cite{kane} are not the focus of our study, they are known to lead to suppression of weak localization \cite{morozov06_short,morpurgo06}, and anomalies
in the density of states \cite{ludwig94_short}.

On the other hand, curvature induced $\pi$-$\sigma$  rehybridzation between nearest neighbors and variations on the next-to-nearest neighbor hopping 
integral $t'$ results in the electrochemical potential (see Fig.~\ref{fig:surface} (c) and the supporting material). In the presence of $t' \approx 0.1$ eV 
the  electronic band structure \cite{reich02_short} is shifted by $3 t'$ and hence, local variations in the hopping due to orbital overlap lead to local changes in chemical potential \cite{thank_steve}. The chemical potential shift due to $t'$ can be written as: 
$\Phi_{{\rm nnn}}(\vec{R}_i) = 6 \sum_{\vec{\delta}} \delta t_i^{'}(\vec{\delta}) e^{-i \vec{\delta} \cdot \vec{K}}$ where $\delta t_i^{'} = (- t'/3+V_{pp\sigma}^{'}/2) [(\vec{\delta} \cdot \nabla) \nabla h]^2$ with \cite{tang} $V_{pp\sigma}^{'} \approx + 1.4$ eV  and 
$\vec{\delta}$ are the next nearest neighbor vectors). The rehybridzation comes from the coupling of $\pi$ orbital to
the three nearest $\sigma$ bonds and antibonds, and to the $3s$ orbitals at three neighboring sites at higher energy  $\epsilon_s^*$ in the presence of the curvature. 
The coupling with the $sp^2$ hybrid on the neighbor is given by $V_{s^*p\sigma}$ and
the bond energies and antibond energies are  $\frac{1}{2}(V_{sp\sigma}^2 +V_{pp\sigma}^2)$.
So in second-order perturbation theory the shiftÊ in each $\pi$ level is given by \cite{thank_walter}
\be
\delta\epsilon_\pi =\left[\frac{1}{2}(V_{sp\sigma}^2 +V_{pp\sigma}^2)\left(\frac{1}{\epsilon_\pi-\epsilon_b}+\frac{1}{\epsilon_\pi - \epsilon_a}\right) +\frac{V_{s^*p\sigma}^2}{\epsilon_\pi-\epsilon_s^*}\right]=-3.23 {{\rm eV}},\ee
 where \cite{walter} $V_{sp\sigma}\!=\! 1.43 {\rm eV}$, $V_{pp\sigma} \!=\! 7.13 {\rm eV}$, $\epsilon_\pi\!=\!-11.07 {\rm eV}$, $\epsilon_b\!=\!-24.34{\rm eV}$, $\epsilon_a\!=\!-3.34{\rm eV}$ and $V_{s^*p\sigma}^2/(\epsilon_\pi\!-\!\epsilon_s^*)\!=\!-1.01{\rm eV}$ . 
Therefore, the combined effect of rehybridization and next-to-nearest neighbor hopping is:
\be
 \Phi({\bf r}) = - \alpha \frac{3a^2}{4}(\nabla^2h)^2,
 \label{eq:Phi}
\ee
where $\alpha = -3 t^{'}+ 9 V_{pp\sigma}^{'}/2 + |\delta \epsilon_{\pi}|=9.23$eV.

 The elastic free energy of the membrane Eq.\eqref{eq:harmonicF} and the effective electronic Hamiltonian Eq.\eqref{eq:DiracH} together with relations between potentials an curvature Eqs.(\ref{eq:A}-\ref{eq:Phi})  define a coupled problem between curvature and fermion bilinears. Of our particular interest is the connection between the local mean curvature couples to the charge density Eq.\eqref{eq:Phi}. Hence, the charge neutrality point will be off-set by $\mu_0\!>\!0$ in the presence of curvature. Further, a flatter region with $|\nabla^2h|\!<\!1/R_c$ will be locally electron doped while a bumpier region with $|\nabla^2h|\!>\!R_c$ will be locally hole doped with a critical length scale $R_c\!=\!\sqrt{3 \alpha/4\mu_0}a$. 
 In turn, spatially varying charge density $n({\bf r})$ enhances rippling tendency  by renormalizing the bending rigidity 
 \be
 \kappa({\bf r}) = \kappa_0- \alpha \frac{3}{4}a^2n({\bf r}).
 \ee
While the curvature can induce electron-electron interaction and electron self-energy corrections, which can be understood via standard perturbation theory starting from the bilinear coupling in Eq. \eqref{eq:DiracH}, these are effects of higher order (fourth order) in weak long wavelength variation of curvature for a nearly flat membrane.  
Another subleading effect of mean curvature is through the ``vector potential'' ${\vec{\cal A}}$ \cite{morozov06_short}. Since the ``field'' $\nabla\times{\vec {\cal A}}$ reverses its direction at the other Dirac point ${\vec K}'$,  in the presence of valley degeneracy, the lowest order effect  on fermion properties of the curvature through  ${\vec{\cal A}}$  is  also of fourth order in curvature\cite{morpurgo06}.  The Gaussian curvature, whose contribution to the bending energy averages to zero (see discussion following Eq.\eqref{eq:bending}), can have more direct consequence in introducing effective random gauge field. The main new result of the present paper is to point out that the leading order effect of the mean curvature is to generate electrochemical potential variation through curvature dependent $\pi$-$\sigma$ rehybridzation. Such electrochemical potential variation will cause charge inhomogeneity, which in turn enhances rippling. 

\begin{figure}[b]
\psfrag{heiginang}{\footnotesize$ h$[nm]}
\psfrag{r}{\footnotesize$r$ [nm]}
\psfrag{VofrinmeV}{\footnotesize$|V|$[meV]}
\subfigure[]{\includegraphics[scale=0.43]{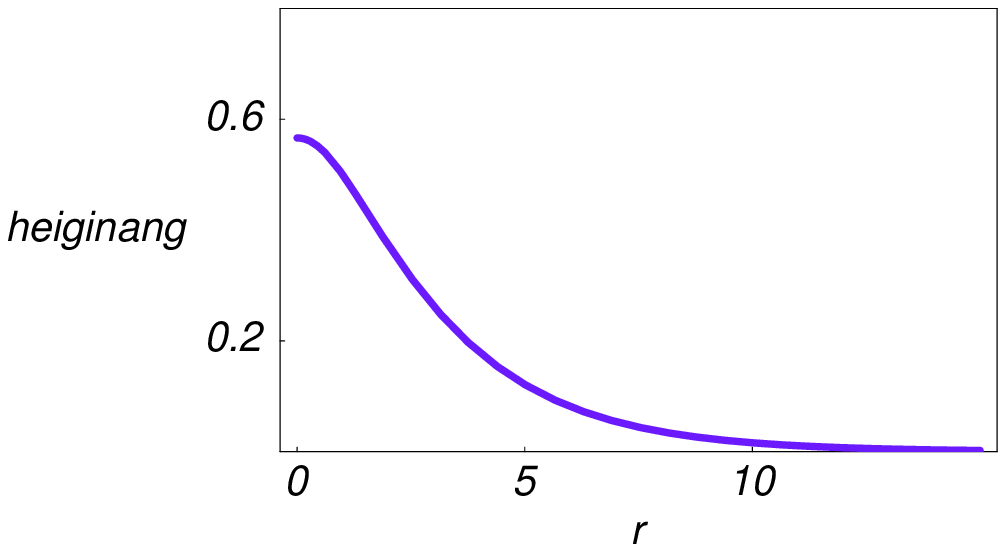}}
\subfigure[]{\includegraphics[scale=0.43]{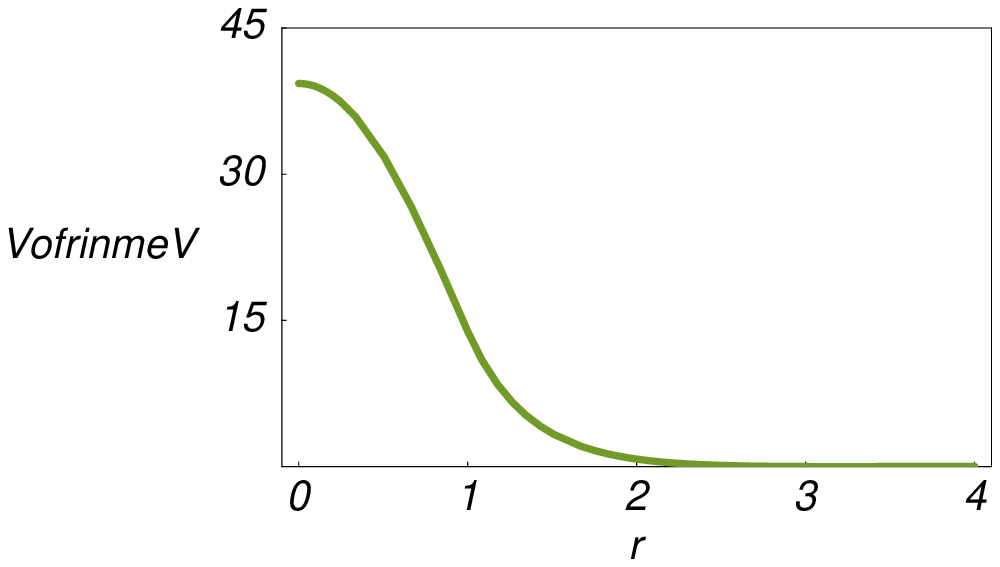}}
\subfigure[]{\includegraphics[scale=0.43]{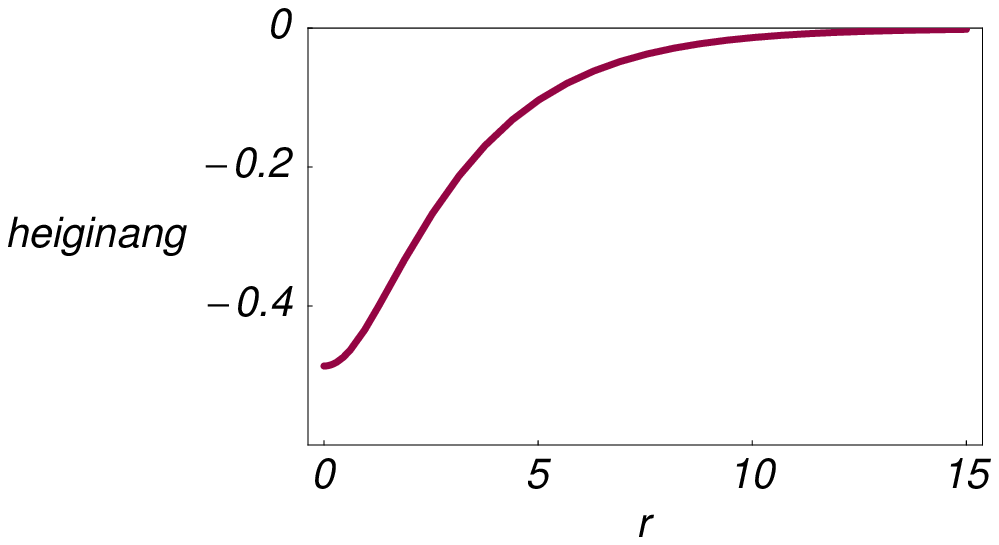}}
\subfigure[]{\includegraphics[scale=0.43]{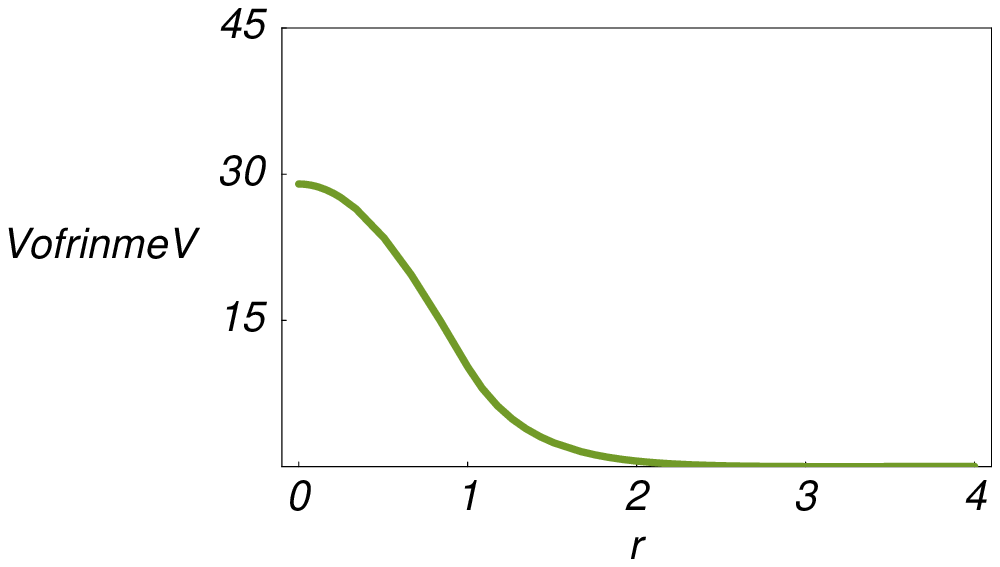}}
\caption{{\small Configurations for ripples 
due to (a-b) a pillar of height $l\!=\!0.5$nm and (c-d) a pit of depth $l'\!=\!9.2$nm, both cases with the radius $R\!=\!1$nm. The membrane configurations $h(r)$ in (a) and (c)  are solutions of Eq.(\eqref{eq:MFh}) for $\xi=1.3{\rm nm}$ and $\xi_\kappa\!=\!2{\rm nm}$(see supporting material). The corresponding electro chemical potential  $V(r)$ are shown in (b) and (d) as a function of $r$.}}
\label{fig:ripple}
\end{figure}

 In principle, the coupled problem defined by Eqs.(\ref{eq:harmonicF}-\ref{eq:DiracH}) should be solved self-consistently for a given configuration of external potential $V({\bf r})$. However, when thermal fluctuation of the membrane is weak, the mean-field theory treatment of the membrane, i.e. solving the Euler-Lagrange equation for the equilibrium configuration of $h({\bf r})$ that minimizes the free energy Eq.~\eqref{eq:harmonicF}, allows tractable analytic progress\cite{swain99}.  
The Euler-Lagrange equation of interest is
\begin{equation}
(\kappa \nabla^4-\gamma\nabla^2+v)\delta h({\bf r})=vs({\bf r})
\label{eq:MFh}
\end{equation}
where $\delta h\equiv h({\bf r})-h_0$. Eq.\eqref{eq:MFh} defines two relevant length scales: $\xi_{\kappa}\!\equiv\!(\kappa/v)^{1/4}$ and  $\xi \!\equiv\! (\kappa/\gamma)^{1/2}$. $\xi_{\kappa}$ determines how closely the membrane follows the external potential and $\xi$ determines the spatial extent of curvature variation. While it is hard to measure individual parameters of Eq. \eqref{eq:MFh}, these length scales can be estimated from the data analysis of recent experiments \cite{kim_STM,Ishigami_STM}. In particular, Ishigami {\it et al.} reported \cite{Ishigami_STM} the graphene height variation under ultra high vacuum condition to range over $\Delta h_{\rm max}\!\sim\!\pm 0.5$nm around the average height of $\bar{h}\!\sim\!0.4$nm with standard deviation of $\overline{\delta h}\!\sim\!0.19$nm. They further analyzed the height-height correlation function  to find the roughness exponent 
to be $2H\!\sim\!1$, which is close to that of SiO$_2$ substrate. They also find the roll over length scale (the length scale at which the correlation rolls over to saturated behavior) to be larger ($\sim 32$ nm) for graphene than for SiO$_2$ ($\sim 23$ nm). This analysis suggests that graphene is following the substrate potential in a coarse grained and smooth manner. 
We construct a model description  with adjustable length scales for a graphene membrane on a random supporting potential. We then determine the length scales guided by observations of Ref.\cite{Ishigami_STM}. From this model we calculate the associated electrochemical potential variation which can be compared with future measurements.

We  model graphene configuration by investigating the solution of Eq. \eqref{eq:MFh} for a piecewise constant potential. As a specific example, we considered a pillar of radius $R$, height $l$ (i.e. $s(r)\!=\!l$ for $\rho\!\leq\! R$ and $s(r)\!=\!0$ for $r\!>\!R$) and a pit of radius $R$ and depth $l'$ ($s(r)\!=\!-l'$ for $r\leq R$ and $s(r)\!=\!0$ for $r\!>\!R$). 
We chose $l\!=\!0.5$nm, $l'\!=\!9.2$nm, $R\!=\!1$nm to generate single ripple configuration with the height variation range of $\Delta h_{\rm max}\!\sim\!\pm 0.5$nm extending over $\Delta r\sim10$nm radius (see Fig.\ref{fig:ripple}) as observed in Refs.\cite{Ishigami_STM,kim_STM,meyer07_short} for $\xi\!=\!1.3{\rm nm}$ and $\xi_\kappa\!=\!2{\rm nm}$. The specific values of $\xi$ and $\xi_\kappa$ were determined based on the observed values of $\overline{\delta h}$, the roughness saturation scale and the extent of ripple $\Delta r$.
Fig.\ref{fig:ripple} shows configuration of a single ripple as a solution to Eq. (\ref{eq:MFh}) (see the supporting material for details) for such potential and corresponding electrochemical potential Eq. (\ref{eq:Phi}). The main feature is that despite the abrupt nature of the supporting potential, the membrane shows extended ripple configuration. Meanwhile, the electrochemical potential variation associated with the curvature has shorter range determined by $\xi_\kappa$.
Solution is qualitatively similar for different piecewise constant supporting potential such as step like potential. 
 
\begin{figure}[b]
\psfrag{x}{$x$\,[nm]}
\psfrag{y}{$y$\,[nm]}
\psfrag{h}{\footnotesize$\delta h$\,[nm]}
\psfrag{0}{\footnotesize$0$}
\psfrag{1}{\footnotesize$1$}
\psfrag{10}{\footnotesize$30$}
\psfrag{20}{\footnotesize$20$}
\psfrag{30}{\footnotesize$30$}
\psfrag{-}{\tiny$-$}
\psfrag{meV}{\footnotesize meV}
\psfrag{n10}{\footnotesize$-30$}
\psfrag{n15}{\footnotesize$-1$nm}
\psfrag{15}{\footnotesize$1$nm}
\psfrag{x}{\footnotesize$x$\,[nm]}
\psfrag{y}{\footnotesize$y$}
\parbox[b]{.5\textwidth}{
\subfigure[Corrugated graphene]{\includegraphics[width=.5\textwidth]{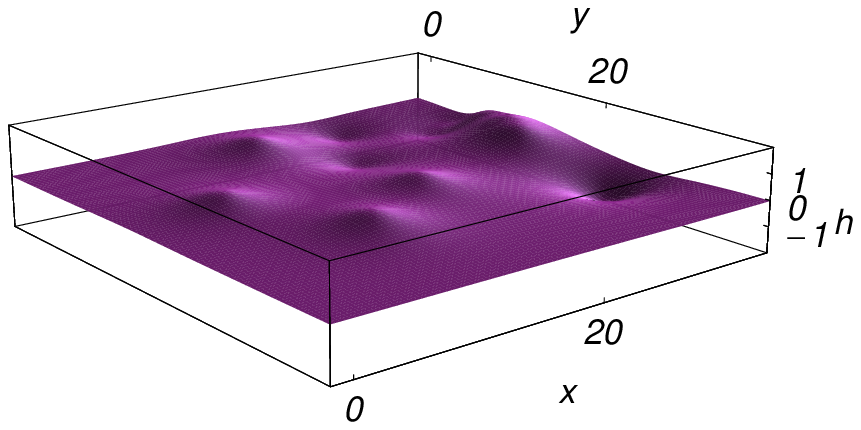}}
}
\parbox[b]{.4\textwidth}{
\subfigure[Statistics on the height variation]{\includegraphics[height=0.2\textheight]
{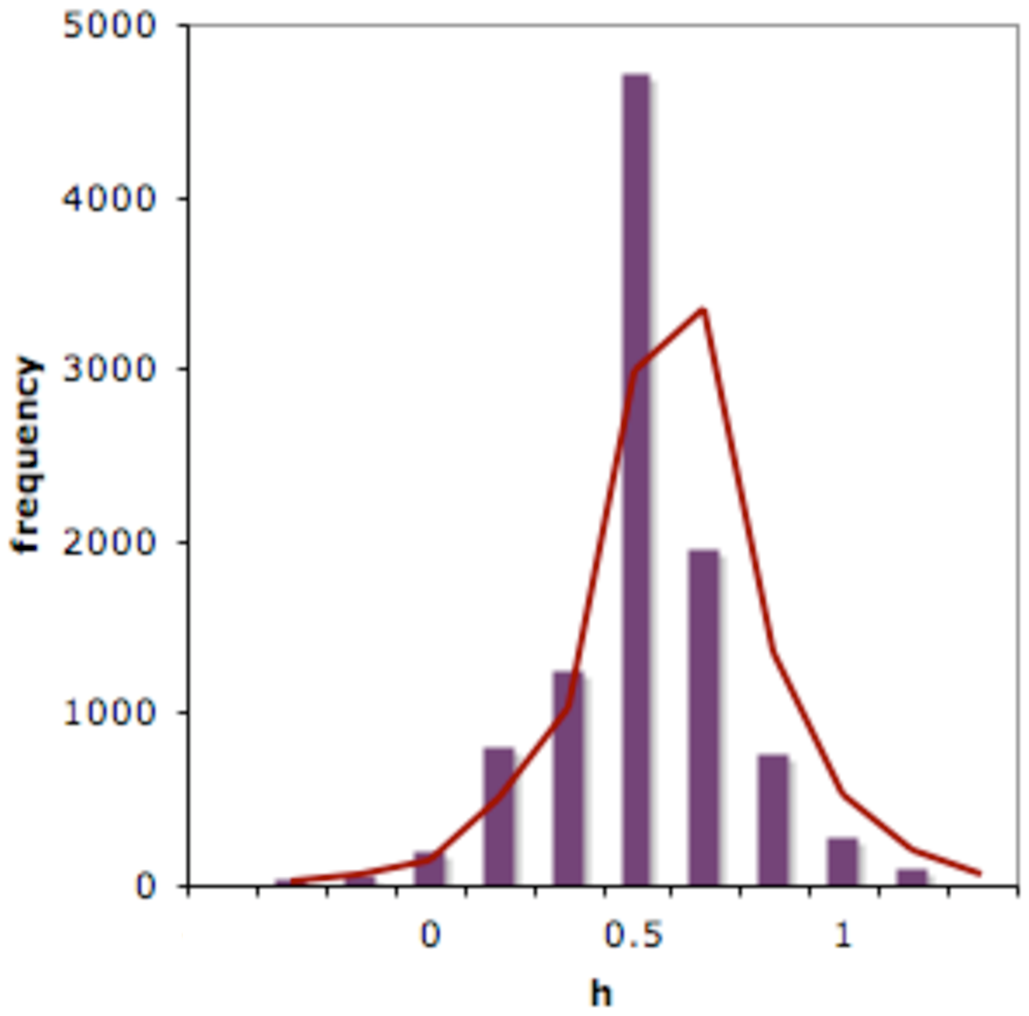}}}
\\
\subfigure[$h(x,y)$]{\includegraphics[width=.35\textwidth]{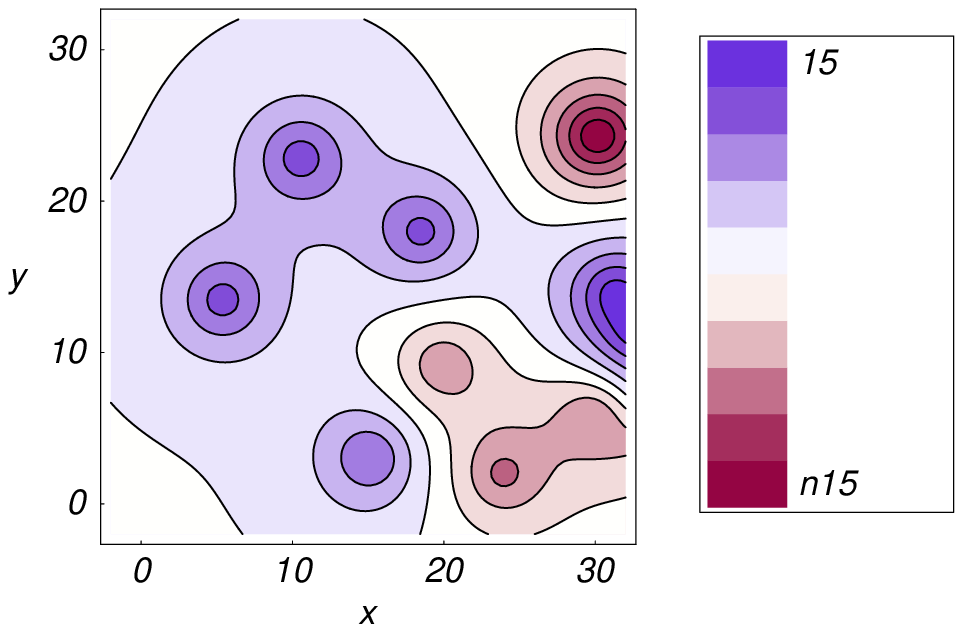}}\hspace{15mm}
\subfigure[$V(x,y)$]{\includegraphics[width=.35\textwidth]{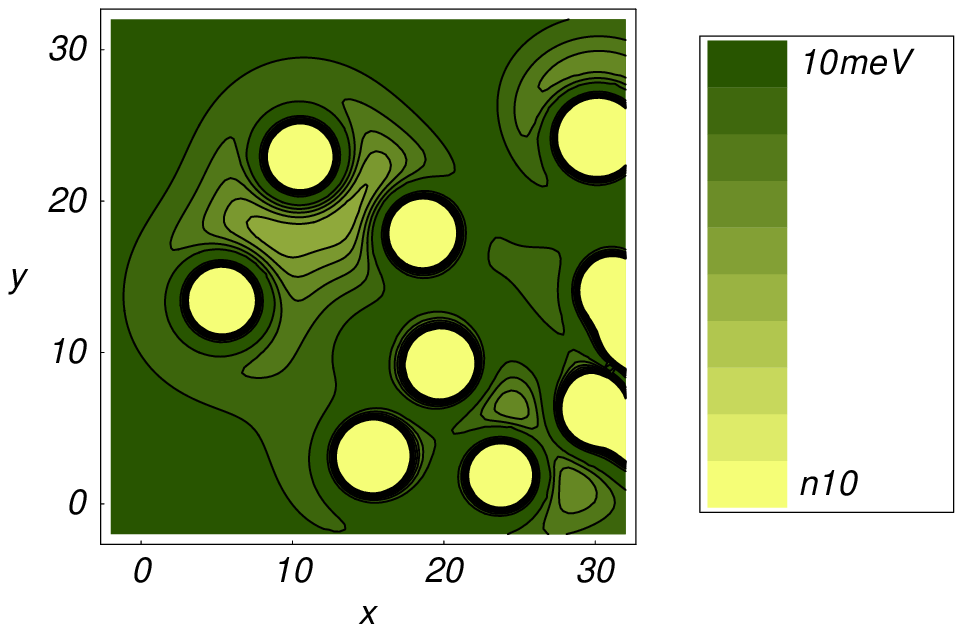}}
\caption{\label{contour}{\small A graphene sheet of size $30$nm$\times30$nm, laid on top of a random supporting potential. 
(a) The model graphene structure showing rippling via spatial fluctuation in the height $h(x,y)$. (b) Histogram showing the distribution of graphene height [nm]. 
(c) Contour plots for the local heigt variation  $h(x,y)$ and (d) the local electrochemical potential variation $V(x,y)$.}}
\end{figure}

Once we understand individual ripple solution, we model the membrane configuration supported by randomly placed supporting potential by randomly distributing supporting centers. In this process, the average distance between the supporting centers are determined based on the observation of ~$20$nm saturation scale for height correlation of SiO$_2$ substrate. Fig.\ref{contour}(a) shows such model configuration.  Fig.\ref{contour}(b) shows the histogram of graphene height distribution. The standard deviation for the distribution is $\overline{\delta h}=0.22$nm close to the value observed in Refs.\cite{kim_STM, Ishigami_STM}, and the maximum deviation from the average height is $\Delta_{\rm max} h= \sim\pm 0.5$nm. This histogram shows that our model configuration captures the height distribution quite similar to that reported in Refs.\cite{kim_STM,Ishigami_STM}. We show the contour plot of height distribution in Fig.\ref{contour}(c) and calculated electrochemical potential in Fig.\ref{contour}(d). The last two figures show that both hills and valleys of graphene correspond to regions of negative electrochemical potential.  

In summary, we investigated the interplay between the membrane aspect and electronic aspect of graphene, which is a unique example of an electronic membrane. We find one-to-one correspondence between the local mean curvature and the electrochemical potential. The spatial variation of mean curvature in the presence of rippling, generates electrochemical potential variation on the graphene sheet itself through $\pi$-$\sigma$ rehybridzation and locally breaks particle-hole symmetry. Furthermore, charge inhomogeneity in turn stabilizes rippling by renormalizing the bending rigidity of the membrane.
We presented a low energy effective theory based on microscopic considerations. We further constructed a model configuration guided by experimentally observed length scales and  found that the observed scale of ripple configuration generates electrochemical potential fluctuation of order $\pm 30$meV. Given that this chemical potential variation is occurring directly on the graphene sheet itself, our calculation suggests that rippling can play a significant role in driving charge inhomogeneity. 
Through our simple modeling we demonstrated how to construct a model configuration based on experimentally observable quantities. 
Our effective theory can be used for more detailed modeling based on detailed understanding of the supporting potential for characterization and design of graphene based electronic elements.

We thank M.~Freedman, A.~Geim, F.~Guinea, W.~Harrison, H.~Liang, M.~Katsnelson, 
S.~A.~Kivelson, J.~L. dos Santos, R.~Shankar, S.-W. Tsai, and A.~Yacoby
for many illuminating discussions. We thank KITP Santa Barbara (NSF Grant
PHY99-07949) for its hospitality during the completion of this work.
A.H.C.N. was supported through NSF grant DMR-0343790.

\end{document}